# Transverse mode selective laser with gain regulation by a digital micromirror device

Zilong Zhang,[1,2,3,*] Yuan Gao,[1,2,3] Xin Wang,[1,2,3] Suyi Zhao,[1,2,3] Yuchen Jie,[1,2,3] and Changming Zhao[1,2,3]

[1] *School of Optics and Photonics, Beijing Institute of Technology, 5 South Zhongguancun Street, 100081 Beijing, China*
[2] *Key Laboratory of Photoelectronic Imaging Technology and System, Ministry of Education of People's Republic of China*
[3] *Key Laboratory of Photonics Information Technology, Ministry of Industry and Information Technology*
*Corresponding author: zlzhang@bit.edu.cn*

**A transverse mode selective laser system with gain regulation by a digital micromirror device (DMD) is presented in this letter. The gain regulation in laser medium is adjusted by the switch of the patterns loaded on DMD. Structured pump beam patterns can be obtained after the reflection of the loaded patterns on DMD, and then it's defocused into a microchip laser medium by a short focal lens, so that the pump patterns can be transferred to the gain medium to regulate the gain distribution. Corresponding structured laser beams can be generated by this laser system. The laser beam pattern can be regulated easily and quickly, by switching the loaded patterns on DMD. Through this method, we show a simple and flexible laser system to generate on-demand laser beam patterns.**

The on-demand generation of laser transverse mode pattern is of great interests in these years, and it has important potential applications in free space laser communication [1-3], optical trapping [4], and 3D optical printing [5,6]. To achieve the selection and adjusting of laser transverse mode, some techniques have been presented, such as the beam phase regulation with cavity mirror [7-9], the beam pattern transformation with optical elements in cavity [10-12], and the intra-cavity loss regulation [13-16] and gain regulation [17-21]. Among these techniques, most of them output changeless laser transverse mode patterns. While, manual operation or elements replacement are asked for the variation of the output beam patterns, which is not flexible enough.

A pattern flexibly and real time regulated laser system is eagerly expected. Andrew Forbes and etc. firstly demonstrated a digital laser for on-demand laser transverse mode generation by using a spatial light modulator (SLM) as an end mirror in laser oscillator to offer the beam phase and intensity of the expected modes [22]. It is able to obtained lots of classical high order transverse modes flexibly by switching the phase patterns loaded on the SLM. Similar works on phase regulation by the SLM in cavity also presented later [23-26]. However, high power or pulsed laser is hardly generated by this mechanism, due to the damage threshold and high loss of the SLM device. The inserted optical wave plate elements in the oscillator are possible to produce complex structured laser beams. While, as the oscillator needs precisely design for a particular plate to generate certain transverse mode, it's not suitable for the generation of multiple on demand laser beam patterns in real time. By the intra-cavity controlling of two-dimensional loss or gain distribution, it's simple to produce structured beams compared with the above methods. As no additional elements are used, the laser cavity can be quite compact, such as the micro cavity, which is of potential for on-chip integration. Nevertheless, for the loss regulation, it is usual to use a defect spot or some circular apertures, and is not easy to obtain multiple shaped loss areas, thus to prevent its ability to produce flexibly varied beam patterns. For the shaping of gain distribution, it's easy to adjust the pump beam patterns to produce a corresponding gain area. Therefore, the output beam patterns will be regulated by the gain distribution. If we can achieve the fast switching of the pump beam patterns, then it's possible to generate fast varied on demand laser beam patterns. Some previous works have used the regulated pump beam patterns for the structured laser beams generation [17-19, 21], it has not referred to a mode selective laser system with digital device controlled pump beam profiles. A recent work by Florian Schepers and etc. proved the possibility of the pump beam shaping by DMD, and high order Hermit-Gaussian (HG) modes were generated by them [27]. However, as a separate cavity mirror was used, the cavity symmetry was required to assist the generation of the modes. And so far, they didn't report the generation of Laguerre-Gaussian (LG), Ince-Gaussian (IG) modes or vortex array beams as they predicted.

In this letter, we demonstrate a transverse mode selective laser system with the gain regulated by a DMD modulated pump beam.

The targeted beam pattern is plotted firstly and then loaded to the DMD system, which reflects the collimated pump beam to a positive lens, focusing the reshaped pump beam to a microchip gain medium and then transferring the originally loaded patterns to it. Under the restraining of the shaped gain area, the output laser beam has a similar profile with the targeted patterns. The selectively generated beams include both HG and LG modes, and the vortex array beams formed by superposition of different HG and LG modes. By the program controlling of the switching of target patterns, on demand real time variation of the structured laser beams can be obtained.

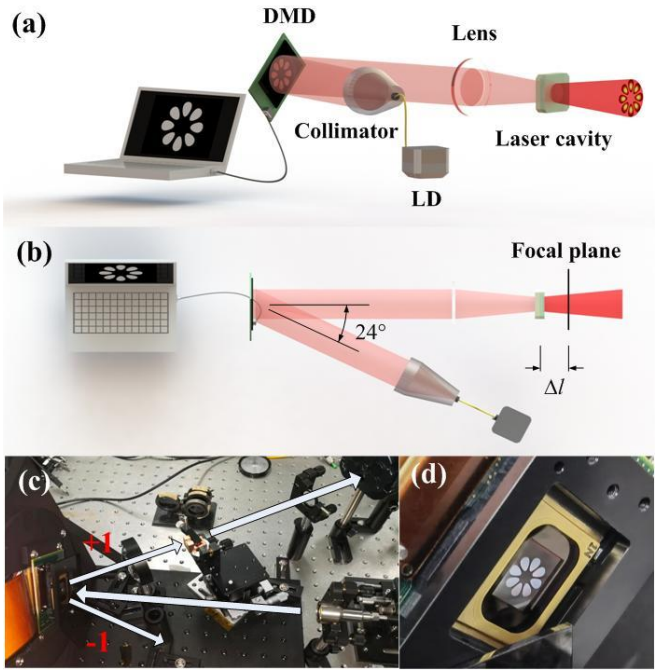

Fig.1 The experimental setup of the transverse mode selective laser system with DMD regulated pump beam.

The experimental setup of this transverse mode selective laser is shown in Fig.1. Fig.1 (a) and (b) are the side view and vertical view of the laser schematic design respectively. Fig.1 (c) is the experimental setup of the laser system, and the DMD with a loaded pattern is shown in Fig.1 (d). A fiber (100 μm core diameter, and 0.22 NA) pigtailed 808 nm diode laser with maximum power of 10 W is used as the pump source. And the output laser beam is collimated by a lens system, to achieve a collimated beam with diameter of about 10 mm. As the beam is collimated pretty well, the beam can maintain about this diameter for several meters. A DMD system is placed on the path of the collimated beam, and it can reflect the pump beam to a focusing lens (focal length f=600 mm), which focuses the reflected pump beam into the microchip laser cavity. If the micro mirror of DMD is placed in the state of reflecting the pump beam into the direction of laser cavity, the state is regarded as '+1', or it will be in the state of '-1'. As the short edge of micromirror array is slightly larger than the diameter of pump beam, the pump power can be reflected and focused into the cavity to the maximum, except for the loss introduced by the protecting glass window, the reflectivity of the mirrors, and the diffracted beams by the grating effect of the mirrors. It's measured that, about a quarter of the pump power can be reflected into the laser cavity with all mirrors in the state of '+1'. For the sake of regulating the pump beam profile, structured patterns should be loaded to the DMD system, leading to the corresponding states of the mirrors switched. Then the loaded patterns can be transferred into the beam cross section. After the focusing of lens, the pump beam profile can maintain the pattern until quite close to the focal plane, where the pump profile is always a circular spot with a diameter of about 270 μm. A 400 μm thick Nd:YAG (stacked with a 1 mm $LiTO_3$) microchip cavity with $5\times5$ mm² cross section is placed in front of the focal plane with an interval of $\Delta l$ =5 mm, where the pump beam size has a maximum diameter of about 900 μm. The location of the micro cavity should satisfy three factors, which are the pump power density for the laser's successful oscillation, the pump beam size for the matching with the aimed transverse modes and the pattern fidelity for the effective selection of transverse modes. The beam waist diameter of the generated $TEM_{00}$ mode are calculated to be 300 μm with a divergence half-angle of 1.1 mrad, and the beam waist diameter for the high order modes are about 520, 670, and 800 μm for the order of 2, 3, and 4, respectively. And these diameters correspond well with the derived ones based on the measured divergence angles of laser beams. The pump beam has an approximately normal incident angle to the microchip cavity, which can be proved by the concentric locations of the output laser and the residual pump beam. In the experiment, we found that slightly changing of the incident angle by tuning the cavity attitude angle has little influence on the output beam pattern. While, the relative pump energy intensity distribution under the same pattern will greatly influence the transverse modes composition of the output beam. As the relative pump energy intensity distribution could be adjusted by slightly tuning the incident angle of the collimated super-Gaussian beam into the DMD surface, it's possible to adjust the selectively generated vortex array laser beams under an ambiguous pump beam profile in cavity.

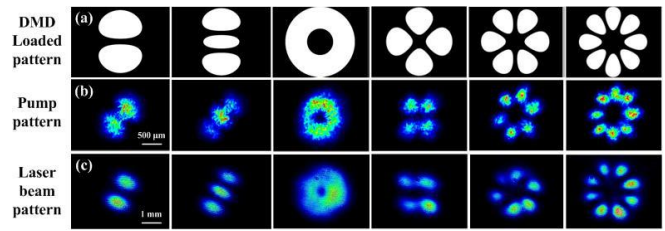

Fig.2. The targeted patterns loaded on DMD and the corresponding pump beam profiles in laser cavity.

The targeted patterns loaded on DMD are shown in Fig.2(a). From the first to the last column of Fig.2(a), the targeted patterns are generated from the transverse modes of $HG_{01}$, $HG_{02}$, $LG_{01}$, $LG_{0,2}+LG_{0,-2}$, $LG_{0,3}+LG_{0,-3}$, and $LG_{0,4}+LG_{0,-4}$. The symbol '+' means a coherent superposition or the phase locking state of two transverse modes. The sizes of the patterns were optimized for a lot of times, also considering the factors of pump power density and pattern fidelity in laser cavity. With these loaded patterns, we can achieve the pump beam profiles in laser cavity as shown in Fig.2(b). It can be seen that the pump beam profiles are basically similar to the originally loaded patterns. Note that, there is a clockwised rotation of 45 degrees due to the 45-degree placement of the DMD. Due to the uniformity of the original pump beam, the diffraction of the micromirror array formed grating effect and the aberrations from the short focal lens, the pump beam pattern on the microchip cavity has some defects. Nevertheless, the easily generated corresponding output beam patterns reveal a strong tendency of the selective effect on the finally generated transverse mode. The selected laser beam patterns by the DMD gain regulation laser system are shown in Fig.2(c). The $HG_{01}$, $HG_{02}$, $LG_{01}$, $HG_{11}$ (or $LG_{0,2}+LG_{0,-2}$), $LG_{0,3}+LG_{0,-3}$, and $LG_{0,4}+LG_{0,-4}$ transverse modes are obtained respectively. For the lower order mode, its loaded pattern can deliver more pump power than the high order one, due to the super-Gaussian beam has a centralized energy distribution, then the total pump power supplied for lower order mode is smaller than the higher mode. Fig.3 shows the pump power conditions for the experimentally obtained modes, with the corresponding loaded patterns on DMD listed at the top of the graph. The histogram presents the pump powers supplied to different modes. The entire height of each orange pillar presents the total pump power from the LD, and the height of the solid area is the corresponding pump power transmitted into the laser cavity with all micromirrors in the state of '+1'. There is a relatively stable ratio between these two heights, and its value is around 55%. And the height of the dark green area is the real pump power transmitted into the microchip cavity corresponding to the certain loaded pattern. And the ratio between the heights of the solid orange area and the dark green area is presented by the blue pentagon symbol, which means how much of the pump beam power is reflected by the micromirrors in '+1' state of a certain pattern. The gray area in the graph give a rough total pump power range for an ordinary gain unregulated laser system to generate corresponding transverse modes by pumping angle tuning method. Comparing the dark green area with the gray area, one can clearly see that the actually pump power required for the gain regulation method is obviously decreased than the unregulated condition.

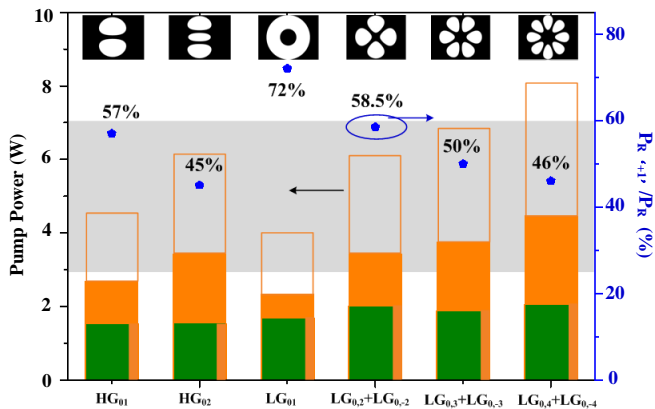

Fig.3 Pump power comparison between the gain regulated and unregulated laser systems. Here, $P_{R'+1'}$ is the pump power transmitted into the laser cavity with a certain loaded pattern, and $P_R$ is that with all micromirrors in '+1' state.

The benefit from the gain regulated laser system is that the particular transverse mode can be easily selected by the gain regulation method with other transverse modes be suppressed. What's more, considering the reflected beam by the mirrors in '-1' state and other losses, actually, the required pump power for particular transverse mode is decreased. Well the output laser power of the selected transverse mode is not obviously affected. Another benefit is that the input power into the microchip cavity is decreased for particular mode, then the thermal effect is also reduced to avoid unnecessary damages. In general, the gain regulation method for the generation of selected transverse modes can achieve a targeted beam pattern easier and require lower pump power in the cavity.

As we know, the laser transverse mode in cavity is not only affected by the gain distribution, but also affected by the phase or the astigmatism of the cavity [28, 29]. The phase is not that easily controlled for a micro Fabry-Perot cavity, while, as the cavity is quite short, the astigmatism can be tuned by the pump energy distribution induced thermal effects. However, under the normal incident pumping angle for a certain pump beam pattern, the astigmatism is not changed largely. And with the help of a particular gain distribution, the transverse mode is sustained in a very small range. Only the frequency degenerated transverse modes can be generated at a certain pump power range. And it's a common phenomenon for the frequency degenerated transverse modes to form a mode locking states, with the locking phase determining the output beam patterns [19, 30-34]. Then the locking phase can be conversely affected by the gain distributions, which can be realized by slightly tuning of the incident angle of the collimated beam on DMD surface. It's found that, the vortex array laser beams in transverse mode locking states are quite easy to be obtained under the gain regulated condition than no gain regulations. And the pump beam profile inside the micro cavity should be an ambiguous one, but not that clear as in Fig.2(b), so that it has no too strong tendency to achieve one certain beam pattern. As shown in Fig.4(a), the pump beam profile was measured at the position of $\Delta l$ =4.5 mm. Compared with the pump beam profile in fifth column of Fig.2(b), the spots of the pump beam are connected to each other. And this condition can lead to the generation of transverse mode locked beams. Fig.4(b1) and (b2) are two generated modes under relatively low total pump power. The corresponding order of the transverse modes can be clearly observed from Fig.4(b1), which is the mode of $HG_{12}$, having the order of N=3. This also agrees with the order of the targeted transverse mode of $LG_{0,3}+LG_{0,-3}$. When the pump power was increased slightly, another mode $HG_{11}$ can start oscillation. And the beam profile changes to a superposition of $HG_{03}$ and $HG_{11}$ modes. Here, the symbol '&' means the incoherent superposition of transverse modes. When we increase the pump power little more again, the frequency degenerated dual-mode locking oscillation can set up. Fig.4 (b3)-(b5) show the transverse mode locked laser beam patterns, with different locking phases. The corresponding simulations to Fig.4(b) are shown in Fig.4(c), and the text at the bottom give the composition of the transverse modes. The simulations have a rotation angle compared with the experimental ones, which have no influence on the similarity of the two results. The phase switching of the locked modes could be realized by the slightly tuning of the incident angle of the collimated beam on the DMD surface.

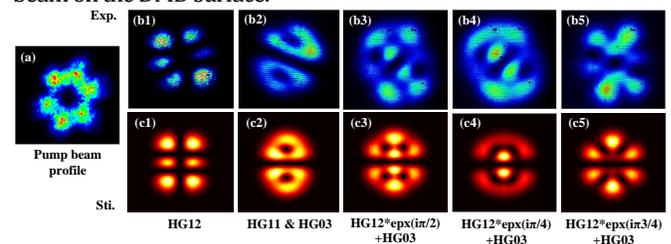

Fig.4 Generation of transverse mode locked beams with an ambiguous pump profile in the microchip laser cavity.

Such conditions with other ambiguous pump beam profiles in laser cavity are also experimentally observed. And various of structured beam patterns in transverse mode locking states can be obtained. The gain regulation is an easier method for the generation of structured laser beams.

However, limited by the maximum power of the pump source we used, the pump beam profile size in the cavity cannot be larger to offer more complex patterns for the definite generation of a certain transverse mode locked beam. If the power is satisfied, it may be possible to definitely obtain a certain complex structured beam pattern in transverse mode locking states or just an incoherent mode superposition condition. And the switching speed of the modes is mainly decided by the reorganization time (tens to hundreds of milliseconds) of the modes in the cavity which is longer than the switching of DMD pattern. Nevertheless, through the experimental results we own now, it already reveals a strong tendency for the transverse modes selected generation with the gain regulation method by a pump beam profile DMD controlled laser system.

In conclusion, we demonstrate a transverse mode selective laser system with the pump beam profile regulated by DMD. Experimentally generated laser beam patterns strongly rely on the regulated pump beam patterns which come from targeted patterns loaded on the DMD system. The gain regulated laser requires much lower pump power than the unregulated one, which is more efficient and not easily disturbed or damaged. With the switch of the loaded patterns on DMD, one can obtain the laser beams on demand easily and quickly.

**Funding.** National Natural Science Foundation of China (NSFC), (61805013).

**Disclosures.** The authors declare no conflict of interest.

**Data Availability.** Data underlying the results presented in this paper are not publicly available at this time but may be obtained from the authors upon reasonable request.